\documentstyle[a4,12pt,epsfig]{article} 
\newcommand {\eettg}      {${\mathrm e^+ e^-}\rightarrow \tau^{+}\tau^{-}\gamma\;$} 
\newcommand {\ttg}        {$\tau^{+}\tau^{-}\gamma\;$} 

\begin{document}
\begin{titlepage}
\begin{center}{\large   EUROPEAN LABORATORY FOR PARTICLE PHYSICS
}\end{center}\bigskip
\begin{flushright}
       CERN-EP/98-033   \\ March 2, 1998
\end{flushright}
\bigskip\bigskip
\bigskip\bigskip\bigskip
\begin{center}{\huge\bf  An upper limit on the anomalous magnetic moment of the \boldmath{$\tau$} lepton 
}\end{center}\bigskip\bigskip
\begin{center}{\LARGE The OPAL Collaboration
}\end{center}\bigskip\bigskip
\bigskip\begin{center}{\large  Abstract}\end{center}
{Using radiative 
${\mathrm Z^0} \rightarrow$ \ttg events collected with the OPAL detector 
at LEP at $\sqrt{s}= M_{{\rm Z}}$ during 
1990--95, a direct study of the electromagnetic current
at the $\tau\gamma$ vertex has been performed  
in terms of the anomalous magnetic form factor $F_2$ of 
the $\tau$ lepton.  
The analysis is based on a data sample of 1429 \eettg events 
which are examined  
for a deviation from the expectation with $F_2 = 0$. 
From the non-observation of anomalous  
\ttg production a limit of $$-0.068 < F_2 < 0.065$$ 
is obtained.
This can also be interpreted as a limit on 
the electric dipole form factor $F_3$ as  
$$-3.8 \times 10^{-16}\, e\,{\rm cm} < eF_3  < 3.6 \times 10^{-16}\, 
e\,{\rm cm} \, .$$ The above ranges are valid at the $95\%$ confidence level.
}
\bigskip\bigskip\bigskip\bigskip
\bigskip\bigskip
\begin{center}{\large
(Submitted to Physics Letters B)
}\end{center}
\end{titlepage}
\begin{center}{\Large        The OPAL Collaboration
}\end{center}\bigskip
\begin{center}{
K.\thinspace Ackerstaff$^{  8}$,
G.\thinspace Alexander$^{ 23}$,
J.\thinspace Allison$^{ 16}$,
N.\thinspace Altekamp$^{  5}$,
K.J.\thinspace Anderson$^{  9}$,
S.\thinspace Anderson$^{ 12}$,
S.\thinspace Arcelli$^{  2}$,
S.\thinspace Asai$^{ 24}$,
S.F.\thinspace Ashby$^{  1}$,
D.\thinspace Axen$^{ 29}$
% ...
}\end{center}\bigskip
\bigskip
$^{  1}$School of Physics and Astronomy, University of Birmingham,
Birmingham B15 2TT, UK
\newline
$^{  2}$Dipartimento di Fisica dell' Universit{\`a} di Bologna and INFN,
I-40126 Bologna, Italy
\newline
% ...
$^{  5}$Cavendish Laboratory, Cambridge CB3 0HE, UK
\newline
% ...
$^{  8}$CERN, European Organisation for Particle Physics,
CH-1211 Geneva 23, Switzerland
\newline
$^{  9}$Enrico Fermi Institute and Department of Physics,
University of Chicago, Chicago IL 60637, USA
\newline
% ...
$^{ 12}$Indiana University, Department of Physics,
Swain Hall West 117, Bloomington IN 47405, USA
\newline
% ...
$^{ 16}$Department of Physics, Schuster Laboratory, The University,
Manchester M13 9PL, UK
\newline
% ...
$^{ 23}$Department of Physics and Astronomy, Tel Aviv University,
Tel Aviv 69978, Israel
\newline
$^{ 24}$International Centre for Elementary Particle Physics and
Department of Physics, University of Tokyo, Tokyo 113, and
Kobe University, Kobe 657, Japan
\newline
% ...
$^{ 29}$University of British Columbia, Department of Physics,
Vancouver BC V6T 1Z1, Canada
\newline
% ...
\newline
%
%
%
%
%
%
%
%
%
%
%
%
%
%
%
%
%
%
%
%
%
%
%
%
%
%
\newpage
\section*{Introduction} \label{intro}

Measurements of the anomalous magnetic moment of the electron \cite{MOMEL} and the muon \cite{MOMMU} by spin precession methods are considered the most precise tests of Quantum Electrodynamics (QED) and are usually expressed in terms of a deviation of their respective $g$-factors from the value of two  
\cite{PDG} 
 
\noindent 
\begin{eqnarray} 
a_{e} & = & \left( \frac{g_e-2}{2} \right) = ( 1159.652193 \pm 0.000010 )
\times 10^{-6} \qquad ,\\ 
a_{\mu} & = & \left( \frac{g_\mu-2}{2} \right) = ( 1165.9230 \pm 0.0084 ) 
\times 10^{-6} \qquad .
\end{eqnarray}     
\noindent 
 
Due to the $\tau$ lepton's short 
lifetime of $(291.0 {\pm} 1.5) \times 10^{-15}\, {\rm s}$, 
its anomalous magnetic moment cannot in practice be measured 
by a spin precession method and 
no direct measurement of $a_{\tau}$ exists so far.  
While  
the hadronic and weak contributions to $a_e$ are very small,
they are no longer negligible for $a_\mu$ and $a_\tau$.
A theoretical prediction for $a_{\tau}$, based purely on QED, 
is $(1173.19 \pm 0.01) \times 10^{-6}$ \cite{samuel}.  
Additional weak and strong contributions \cite{samuel,hamzeh} 
modify this to $(1177.3 \pm 0.3) \times 10^{-6}$. 
Using the total width of ${\rm Z}^0 \rightarrow \tau^+\tau^-$,
ref.\,\cite{masso} indirectly derives an upper limit on
$a_\tau$ of $|a_\tau| < 0.01$ at $95\%$ confidence level.
\begin{eqnarray}
\nonumber
\resizebox{6.0cm}{5.0cm}{%
\epsfig{file=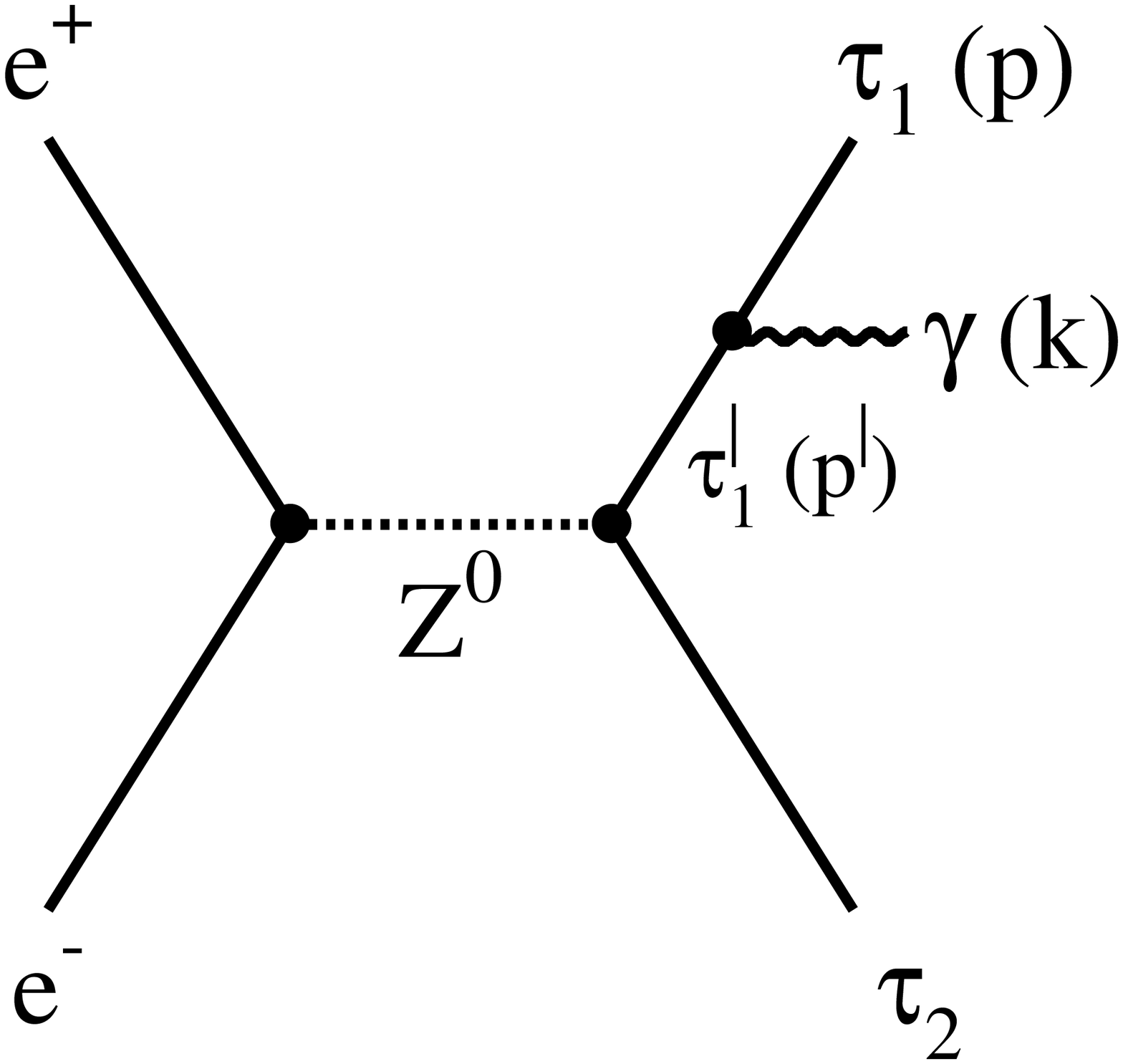}}
\end{eqnarray}
In order to constrain $a_{\tau}$ as suggested by Grifolz and 
Mendez \cite{grifolz}, we have studied the process 
${\mathrm e^+ e^-} \rightarrow \tau^+ \tau^- \gamma$  
in which a final-state photon is radiated from one of the tau leptons, as
shown in the Feynman diagram above. 
The electromagnetic current of a fermion with mass $m$ and charge $e$ can be 
written using the general form factor decomposition 
\begin{eqnarray}\label{current} 
j^\mu_{\rm em} = e{\bar u}(p') \left[ \gamma^\mu F_1(q^2) + \frac{i}{2m} F_2(q^2) \sigma^{\mu\nu} q_\nu + \gamma^5 \sigma^{\mu\nu} q_\nu F_3(q^2)\right] u(p), 
\end{eqnarray} 
with $p',p$ being the four-momenta of the $\tau$ lepton before and after 
the emission of the photon with four-momentum $q$ and 
$q^2 = (p - p^{\prime})^2$.
At $q^2 = 0$,
$F_1(0) = 1$ while $F_2(0) = a_f$, and $e F_3(0) = d_\tau^{\rm el}$ 
define the anomalous magnetic and electric dipole moment, 
respectively. Note that the Standard Model predicts $F_3 = 0$. 

However, this ansatz is not directly applicable to the 
$\tau^\prime \rightarrow \tau \gamma$ vertex in \eettg, since 
the $\tau^\prime$, which emits the photon, is off-shell. Instead, the 
pertinent part of the amplitude must be written as  
\begin{equation} \label{amplitude} 
\frac{i({p'}\hspace{-0.75em}/+m)}{p'^2-m^2}\,ie 
\left[ \gamma^\mu F_1(p^{\prime 2}, q^2)  
       + \frac{i}{2m} F_2(p^{\prime 2}, q^2)\, \sigma^{\mu\nu} q_\nu  
       + \gamma^5 \sigma^{\mu\nu} q_\nu F_3(p^{\prime 2}, q^2)  \right]  
\epsilon(\kappa) u(p) \quad .
\end{equation} 
\hfill\newline
The photon belongs to the final state, so it is real and therefore
$F_2(p^{\prime 2}, q^2)$ is measured at $q^2=0$, but averaged over 
a range of $p^{\prime 2}$ from $m_\tau^2$ to $(M_{\rm Z} - m_\tau)^2$. In this
analysis the minimum value of $p^{\prime 2}$ after the event selection
is  $(13\, {\rm GeV})^2$.  
 
In this paper we search for an excess in the production of 
${\mathrm e^+ e^-} \rightarrow 
{\mathrm \tau^+ \tau^-} \gamma$ events due to a non-vanishing form 
factor $F_2(p^{\prime 2},0)$ as defined by eq.\,(\ref{amplitude}), 
assuming that $F_3=0$. 
Differential photon rates are compared to Monte Carlo predictions for the 
standard $F_1$ and the anomalous $F_2$ term. 
The extracted bound on the number of excess events from the $F_2$ term is 
used to determine an upper limit on $F_2$ averaged over
$p^{\prime 2}$. Henceforth this interpretation of $F_2$ is always implied. 
Conversely, assuming $F_2 = 0$ a limit on $F_3$ is obtained. Because 
the sensitivity of this analysis is not sufficient to measure a value of
$F_2$ as small as predicted \cite{samuel} by the Standard Model (SM), 
the reported 
results mainly address new physics phenomena beyond the SM. 
Such phenomena may occur in the context of composite $\tau$ leptons
\cite{COMPTAU}, leptoquark models \cite{LEPTOQ}, or in models
in which the electroweak symmetry breaking is driven by the 
third quark and lepton generation such as top-condensation 
or top-colour models \cite{TOPCOND}.

It should be noted that the ansatz of eq.\,(\ref{amplitude})
can parametrize modifications of only  
the $\tau^\prime \rightarrow \tau \gamma$ vertex. 
Radiative corrections involving both final-state taus, as well 
as the non-vanishing $p^{\prime 2}$, therefore 
limit a direct interpretation of $F_2$ in terms of the
$\tau$-lepton's anomalous magnetic moment  $a_\tau = F_2(0,0)$.
For physics beyond the SM at an interaction scale
$\Lambda_{\rm new} \gg M_{\rm Z}$, however, there is no such limitation
in the above ansatz. In fact,
as long as $|p^{\prime 2} - m_\tau^2| \ll \Lambda_{\rm new}^2$,
equating $F_2$ with the $a_\tau$ pertaining to the 
new interaction is correct.  
 
The calculation which is used here to predict the distribution of photons 
arising from the different contributions assumes no interference between 
the $F_1$ and the $F_2$ term. The interference term is 
suppressed by $m_\tau^2 / M_{\rm Z}^2$. 
No severe restriction is imposed by this assumption 
for the precision of the $F_2$ measurement described below.
Modifications of the results due to the interference term 
are treated at the end of the paper.

\section*{Monte Carlo simulation} 
 
The Monte Carlo simulation of the process 
${\mathrm e^+ e^-} \rightarrow 
\tau^+ \tau^- \gamma$ with $F_2=F_3=0$ is 
provided by the program KORALZ \cite{KORALZ} including 
initial (ISR) and final (FSR) state photon radiation up 
to ${\cal O}(\alpha^2)$. 
To the extent that the expectation for $F_2$ within the SM is 
small compared to the sensitivity of this analysis, KORALZ is assumed 
to represent the SM expectation throughout this paper.
The $\tau$ decay is simulated by the TAUOLA \cite{tauola} 
package which includes photon radiation from the leptonic decay 
products up to ${\cal O}(\alpha)$ and also from hadronic 
decay products using the program package PHOTOS \cite{PHOTOS}. 
According to studies using the KORALZ MC,
the only source of photons 
contributing to the selected events studied in this analysis will be from 
ISR and FSR. Photons from $\pi^0$ decays do not 
enter as background to this analysis after the event selection.

The contribution of \ttg events coming 
from a non-vanishing form factor $F_2$ is simulated using a calculation 
by Zeppenfeld \cite{ZEPP1} based on the $F_2$ term in eq.\,(\ref{amplitude}), 
assuming $m_\tau = 0$ and neglecting interference.
The resulting differential cross section is given in the Appendix. 
In fact, the approximation of $m_\tau=0$ 
implies a chirality (=helicity) flip in the amplitude 
for the $F_2$ contribution, while the Standard Model radiation always 
conserves chirality. 
As a result, there is no interference between the 
Standard Model and the $F_2$ contribution in the massless limit. 
Conversely, the size of the interference term then checks the
validity of the $m_\tau=0$ approximation. 
A very recent calculation \cite{Gau} of radiative tau pair
production through anomalous electromagnetic couplings 
including interference effects and a finite $\tau$ mass
confirms the validity of the assumptions ($m_\tau = 0$, 
interference neglected) made here. Ref.\,\cite{Gau} 
concludes that anomalous
contributions from initial-state final-state interference, ${\rm Z}^0/\gamma$ interference
and $\gamma$ exchange can also be safely neglected.
 
Events generated from both the $F_1$ bremsstrahlung term (KORALZ) and the 
$F_2$ contribution are processed through a full simulation of 
the OPAL detector \cite{OPALSIM}.  
For the purpose of the efficiency determination for the $F_2$ contribution
(signal), events have been generated by KORALZ and selected 
according to the 5-dimensional differential $F_2$
cross section (see Appendix) employing a `hit or miss' method.
 
\begin{figure}[htbp]
  \begin{tabular}{ll}
  \hbox{\epsfig{file=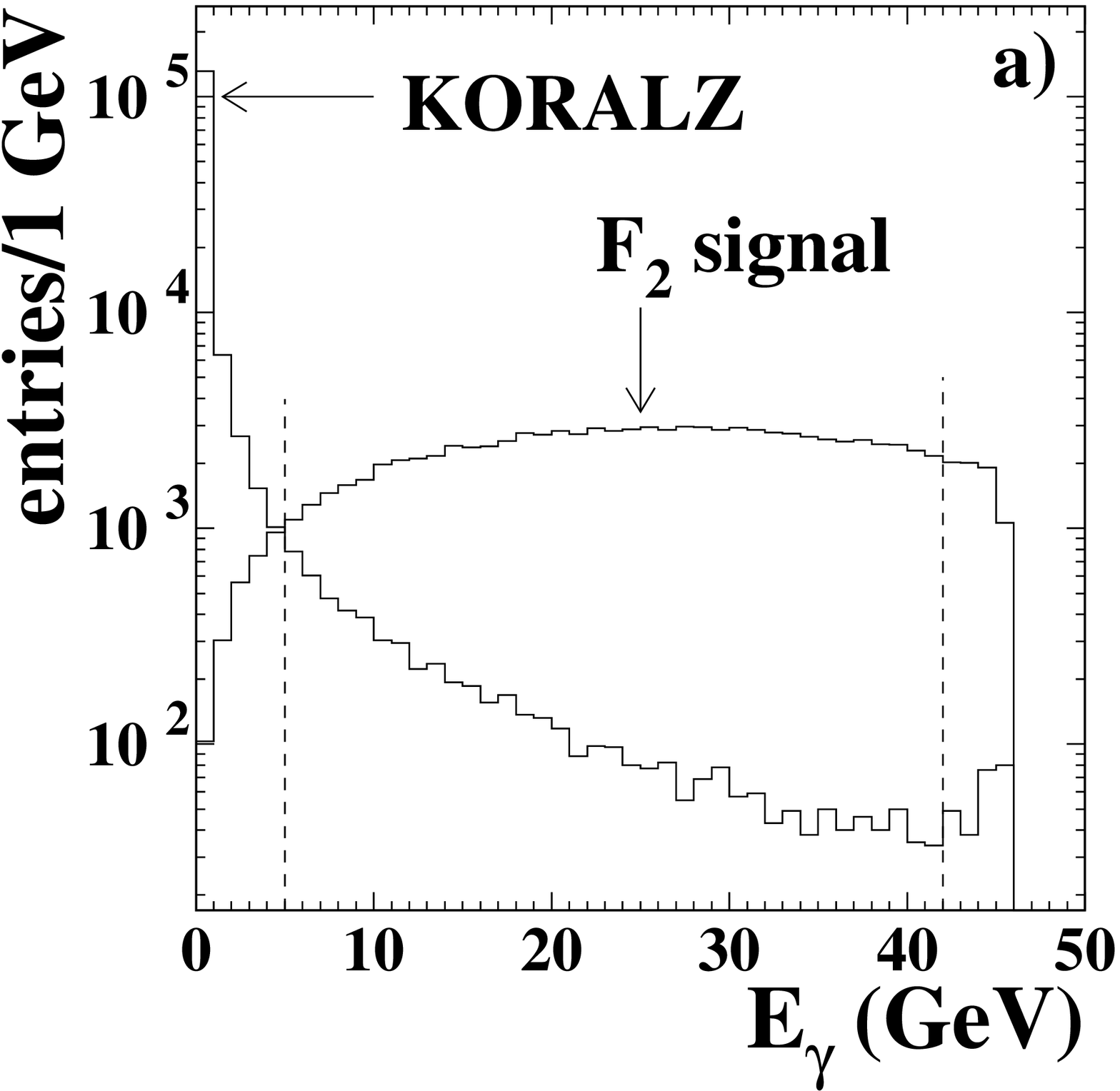,width=0.5\textwidth}}&
  \hbox{\epsfig{file=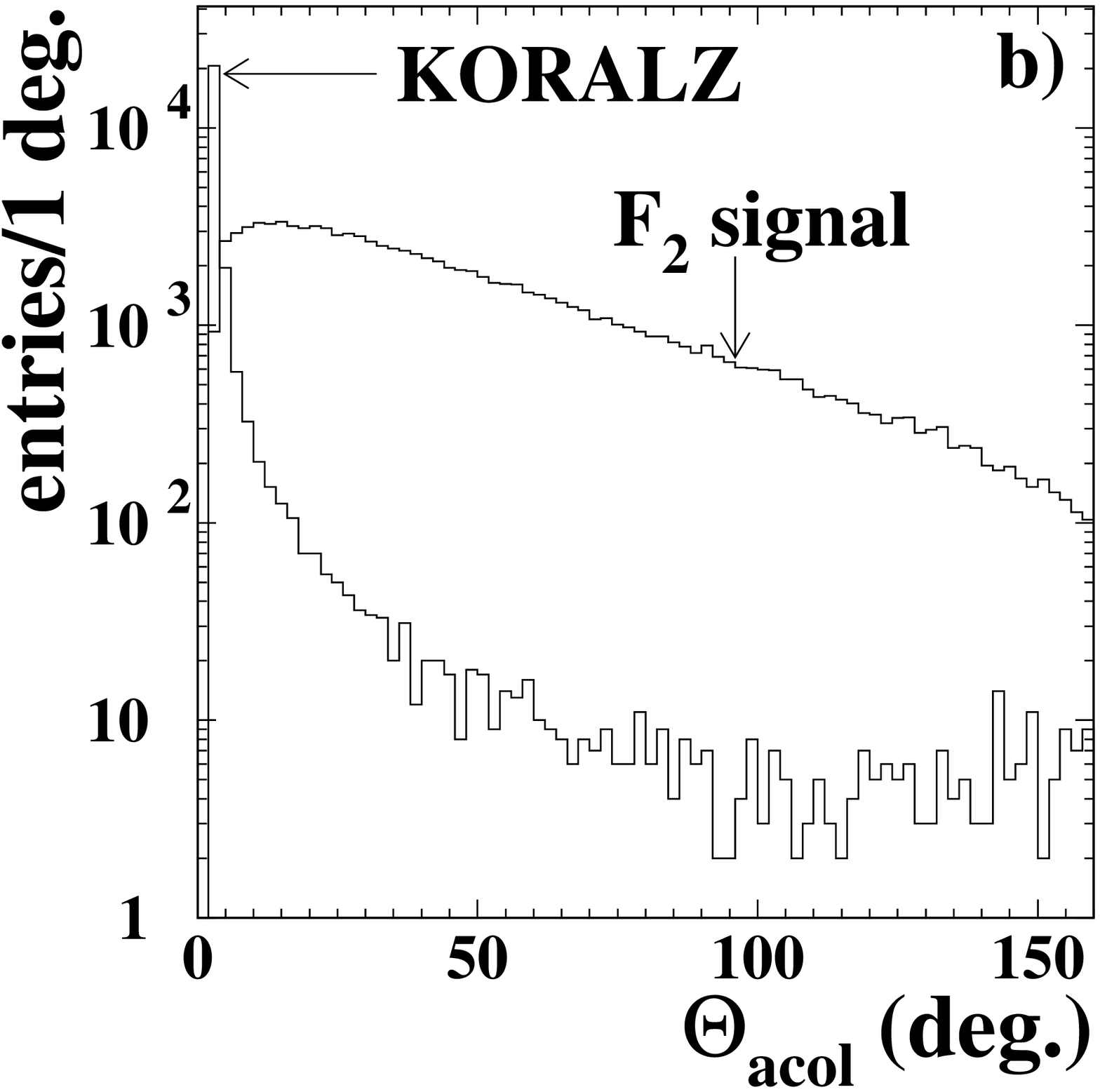,width=0.5\textwidth}} \\
  \hbox{\epsfig{file=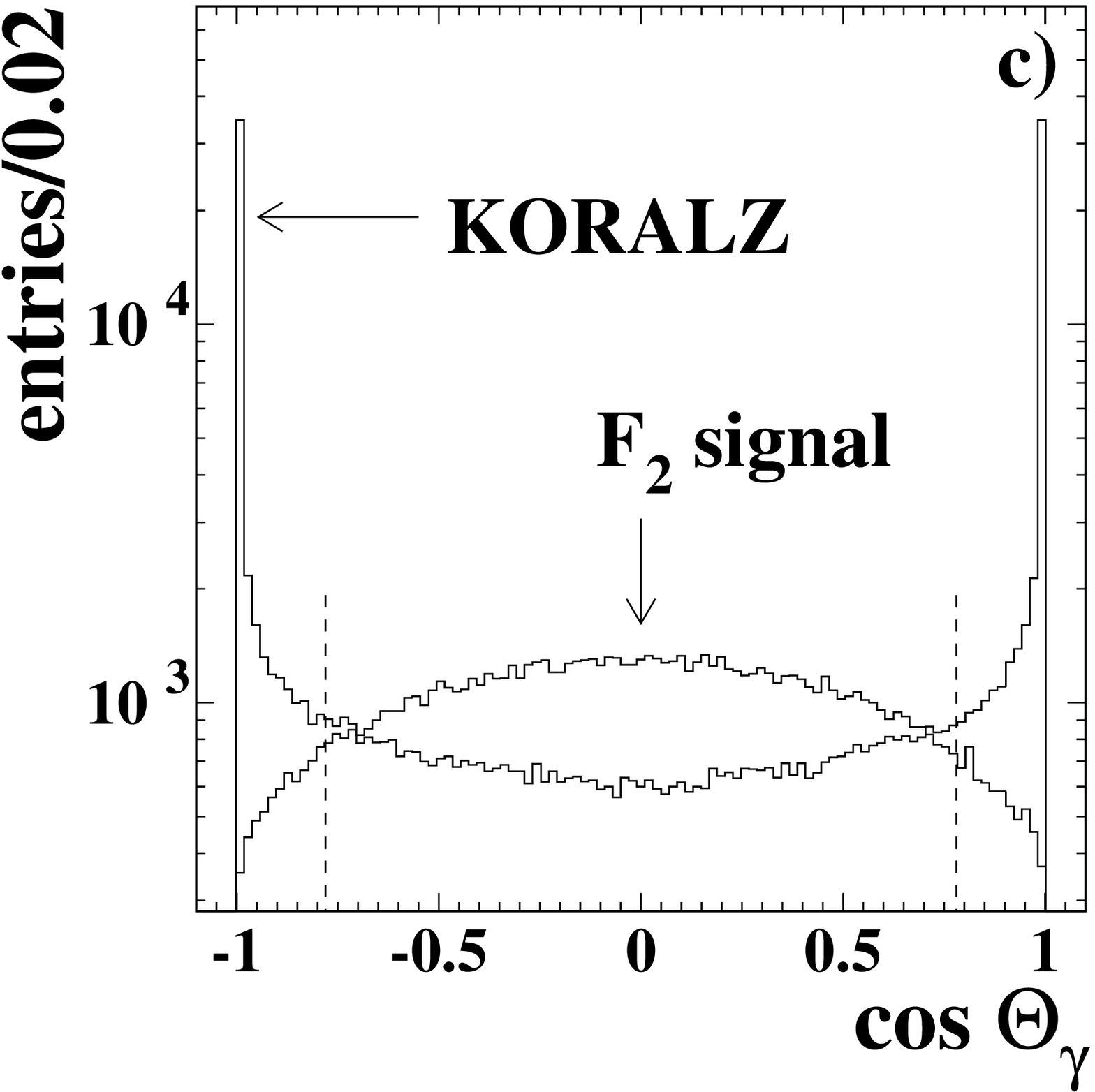,width=0.5\textwidth}}&
  \hbox{\epsfig{file=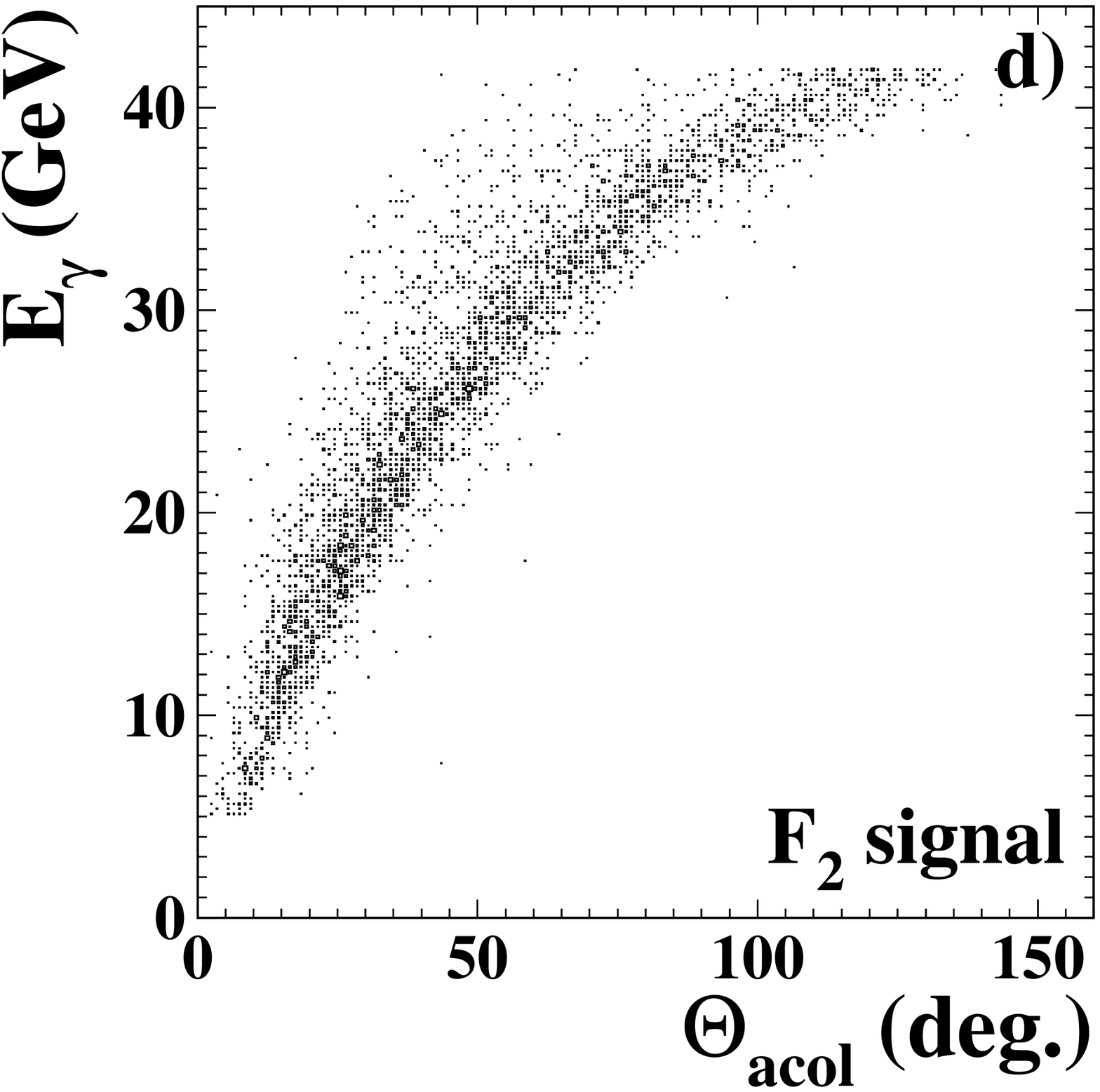,width=0.5\textwidth}}\\
  \end{tabular}
\caption[Signal characteristics]{\it Comparison between the 
$F_2$ signal and the Standard Model expectation represented by 
KORALZ. 
(a) the photon energy $E_\gamma$, 
(b) the acollinearity angle $\theta_{\rm acol}$ of the $\tau$ leptons
and (c) the photon angle $\cos \: \theta_\gamma$ w.r.t. the 
beam direction. The relative normalization of the distributions 
is arbitrary, no detector effects are included. The dashed lines in (a) 
and (c) indicate the acceptance cuts.
(d) photon energy ${E_\gamma}$ vs. acollinearity angle for the $F_2$ signal 
prediction after full detector simulation.
}
\label{signal} 
\end{figure} 
 
Figure\,\ref{signal} shows comparisons of the anomalous contibution and 
the KORALZ prediction in simulated distributions of the energy 
of the radiated photon $E_\gamma$ (a), 
the acollinearity angle $\theta_{\rm acol}$ between the $\tau$ lepton 
directions (b), and the emission angle of the photon with respect 
to the beam direction $\cos \: \theta_\gamma$ (c). Note, that 
the anomalous part is arbitrarily normalized.  
The striking difference between 
the distributions 
suggests that these variables
are useful discriminators for this analysis:
$F_2$ photons appear to be preferentially at high  
energies and are emitted at large angles to both $\tau$'s.  
However, $E_\gamma$ and $\theta_{\rm acol}$
are strongly correlated (fig.\,\ref{signal} (d)), 
and the $\cos \: \theta_\gamma$ distribution is almost
isotropic in an accepted $\theta_\gamma$ range with reduced background
($|\cos \: \theta_\gamma| < 0.78$).   
Consequently, the benefit obtained when using two-dimensional information 
in the above variables has been found to be marginal and also more sensitive to 
systematic effects. Therefore, in what follows
only the photon energy distribution is used.
 
For the simulation of background processes the Monte Carlo 
generators \cite{JETSET} 
JETSET~7.4 (${\rm q{\bar q}}$), RADBAB~2.0 (${\mathrm e^+ e^-}$), KORALZ~4.0 
($\mu^+\mu^-$) and VERMASEREN~1.01 ($2\gamma$) have been used.

\section*{Event Selection} 
 
For this analysis events recorded with the OPAL detector 
during the years 1990 to 1995 at a centre of mass energy $\sqrt{s}=M_{\rm Z}$, 
corresponding to an integrated luminosity of about
$180\,{\rm pb}^{-1}$, have been used. `Off-peak' data were not  
used to avoid deficiencies due to the lack of 
${\rm Z}^0/\gamma$ interference effects in the signal MC.  
The number of produced $\tau$ pairs is about $230\,000$.  
The OPAL detector and its performance have been described elsewhere 
\cite{OPAL}. Isolated final-state photons are detected in the lead glass 
electomagnetic calorimeter covering an angular range in the 
barrel region of $|\cos \: \theta| < 0.81$ with an energy
resolution of $\sigma_E/E \approx 12 \%/\sqrt{E ({\rm GeV})}$. 

In selecting events containing $\tau$ pairs with an additional 
radiated photon, background is expected from ${\mathrm e^+ e^-}$, 
$\mu^+\mu^-$, multihadron, and two-photon events with 
any final state. Lepton pair events are selected by standard 
cuts \cite{tau_selection}
against ${\rm Z}^0 \rightarrow {\rm q \bar{q}}$ (cut on track and cluster
multiplicity), two-photon processes (cut on visible energy) and 
cosmic ray background.
The cut on the acollinearity of the $\tau$ pair is of course omitted
in the preselection since it would also reject most of the 
signal events.  
Then ${\mathrm e^+ e^-}$ and $\mu^+\mu^-$ events are recognized and rejected 
by high detected energy in the electromagnetic calorimeter (ECAL) 
or by high momentum tracks, low energy deposits in the calorimeters and 
$\mu$ chamber hits, respectively. 
The observed $\tau$ decay products are 
required to lie in a cone of half opening angle of $35^\circ$. 
We assume the $\tau$ flight direction to be identical with the  
cone axis, defined by the vector sum of the associated tracks and 
all neutral clusters.
A \ttg candidate event is selected if a
photon candidate of at least $5$~GeV and less 
than $42$~GeV energy deposit is found outside both cones. 
The high energy cut is imposed to  
avoid the energy region where the $m_\tau = 0$ 
assumption in the $F_2$ MC has an impact on the $E_\gamma$ distribution.  
To avoid losing $\tau$ decay products inside the beam pipe,  
$|\cos \: \theta_{\tau}| < 0.9$ has to be valid for both $\tau$ cones. 
A sample of $3\,435$ events survive this preselection.  
    
Background from non-$\tau$ events is further reduced by the following 
requirements: 
\begin{itemize} 
\item
to suppress initial state radiation
the photon has to be in the barrel region of the detector 
($|\cos \: \theta_\gamma | < 0.78$). 
 
\item
to reject ${\mathrm e^+ e^-}$ and $\mu^+\mu^-$ events, the visible energy  
or visible momentum of the more energetic $\tau$ candidate must be 
smaller than $35 \, {\rm GeV}$. 
 
\item
the scalar sum of the momenta of the detected decay products 
of both tau candidates 
and the photon must be smaller than $75  \, {\rm GeV}$;
furthermore events for which both $\tau$ cones are
identified as $\tau \rightarrow \mu \nu {\bar \nu}$ decays are 
rejected. These cuts add to the suppression of $\mu^+\mu^-$ events. 
 
\item
only events with 2 or 4 charged tracks (1--1 and 1--3 topologies) 
are retained. 
 
\end{itemize}  
 
The three-body final state of the signal process 
${\mathrm e^+ e^-} \rightarrow \tau^+ \tau^- \gamma$ is  
completely determined by two independent variables, i.e. 
the acollinearity angle  
between the $\tau$ leptons can be  
calculated from the measured photon energy $E_\gamma$ and the measured angle  
between the photon and the $\tau^-$. 
The measured acollinearity angle is required 
to agree within $\pm 50^\circ$ with the calculated angle. 
This cut greatly reduces multihadron background, two-photon events 
and incorrectly reconstructed events.  
 
The above selection results in a total of 
$1429$ ${\mathrm e^+ e^-} \rightarrow \tau^+ \tau^- \gamma$ candidate 
events. 
The contribution of background 
from ${\mathrm e^+ e^-}$, $\mu^+\mu^-$ and multihadron events to this sample is
estimated to total $(0.13 \pm 0.13)\%$. 

\begin{figure}[htb] 
\centerline{\epsfig{file=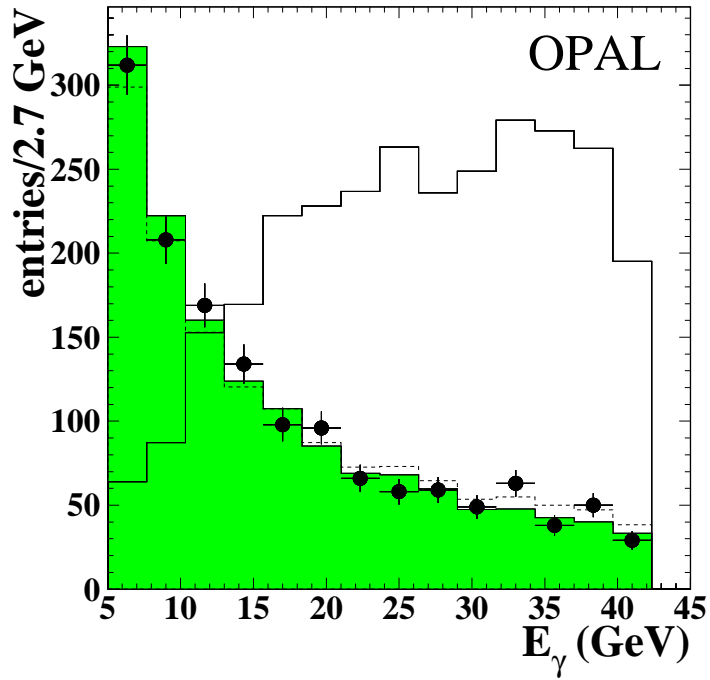,width=100mm}} 
{\caption{\label{egamma}
{\it  Photon energy spectra for data (points), 
KORALZ (shaded histogram, normalized to the data), 
and signal (open histogram, arbitrarily normalized). 
The dashed line shows the Monte Carlo prediction for $F_2 = 0.064$, also 
normalized to the data. }}}
\end{figure}

Figure\,\ref{egamma} shows the observed 
distribution of the measured photon energy for the  
selected events. 
Superimposed are the expectation from the Standard Model (KORALZ), 
normalized to the number of data events (shaded) and 
the distribution of $F_2$-produced photons including 
full detector simulation (open histogram). 
In order to extract a limit on the $F_2$ form factor the data distribution 
of fig.\,\ref{egamma} is fitted to a sum of both MC contributions using a 
binned likelihood ${\cal L}$ assuming a Poisson distribution of the data events.
%
%
%
%
%
%
In this fit the sum of both contributions has been normalized 
to the number of observed data events for each assumed value
of $F_2$. 
To test the method, we have performed fits to 
Monte Carlo event samples of the size of the data sample,
using a $5 \%$ and a $10 \%$ $F_2$ contribution, respectively. In both 
cases the input values were reproduced ($0.045 \pm 0.010$ and
$0.096 \pm 0.007$, respectively).

\section*{Fit Results} 

\begin{figure}[htbp] 
\resizebox{!}{!}{%
\epsfig{file=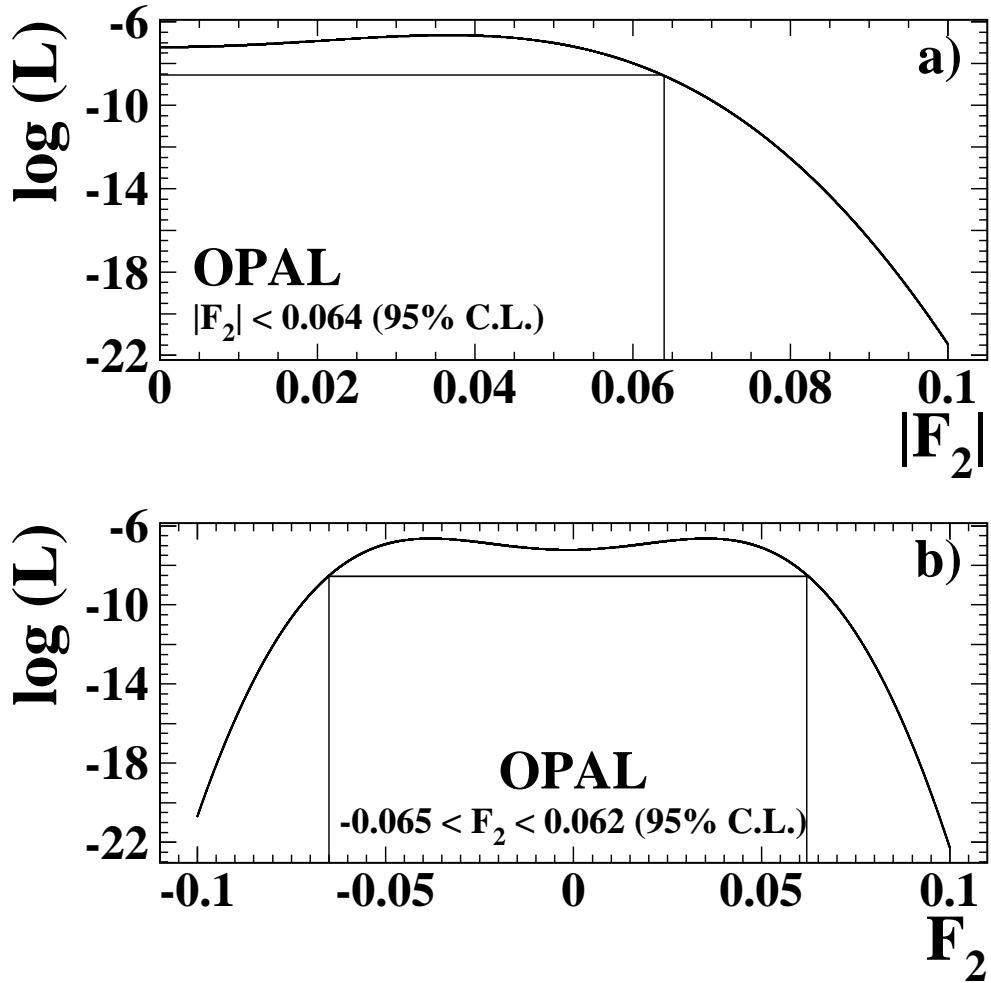,width=\textwidth}} 
{\caption{\label{likeli}
{\it (a) Likelihood as a function of the fit parameter $|F_2|$. 
     (b) Likelihood as a function of $F_2$ taking into account the
         interference term as described in the corresponding section 
         of the text.}}} 
\end{figure}

Figure\,\ref{likeli}(a) shows the dependence of the likelihood 
function on $|F_2|$. 
The most probable value is $|F_2| = 0.037$ which is offset from, but 
consistent with zero within about one standard deviation 
($+0.015 \atop -0.028$) which is
evident from the shallowness of the maximum of ${\rm log} {\cal L}$ in 
fig.\,\ref{likeli}(a).
The $95\%$ confidence level value is obtained from fig.\,\ref{likeli}(a) 
at the point where ${\rm log} {\cal L}$ has dropped by $1.92$ units from its
maximum as

\begin{equation} 
|F_2|  < 0.064  \qquad {\rm at}  \qquad 95 \% \, {\rm C.L.} 
\end{equation} 

The analysis has also been performed by normalizing the KORALZ MC 
to the integrated luminosity of the data. In this case 
the luminosity has been inferred from the total number of 
$\tau$-pair events using the standard OPAL $\tau$-pair selection.
While intuitively one might expect a tighter constraint on $|F_2|$,   
the necessity to know the detection efficiency introduces an
additional uncertainty not present in the approach described 
above which is only sensitive to a difference in shape of the 
$E_\gamma$ distribution between data and MC.
Both effects have been tested (see section on systematic errors).
When normalizing to the integrated luminosity the uncertainty  
of $\pm 6\%$ in the knowledge of the detection efficiency
yields a limit on $|F_2|$ even slightly higher ($0.065$) 
than that obtained using the shape information only. 

\section*{Systematic Errors} 
 
The systematic uncertainties due to the selection cuts, 
the photon detection efficiency, non-$\tau$ background, 
binning effects, Monte Carlo statistics and normalization, 
and the calibration uncertainty in the photon energy measurement
have been studied. 
The omission of the interference term is discussed separately below. 
 
Variations of the {\it selection cuts}  
indicate a systematic uncertainty on 
the limit on $|F_2|$ of about $0.005$, 
the largest influence coming 
from varying the visible energy cut for the $\tau$ cone 
from $35$~GeV to $32$~GeV. 
 
The effect of {\it binning} has been studied by varying the number of 
bins in the energy spectrum from 15 to 20  
and by moving the bin boundaries by half a bin width.  
This leads to a maximum change in the limit on $|F_2|$ of 
$+ 0.002$.  
 
The {\it photon energy calibration} 
has been investigated as a source of systematic error. 
The agreement in the energy measurement of the 
electromagnetic calorimeter between data and MC is better than
$0.9 \%$, determined from a comparison of $\pi^0$ invariant masses
involving all photon energies.
A systematic shift of the photon energy by this amount results in 
a systematic uncertainty on the limit on $F_2$ smaller than $0.001$. 

The uncertainty in the 
{\it description of the photon detection efficiency}
and its dependence on the 
energy can make an important 
contribution to the systematic error of this analysis. 
An imperfect description of the efficiency would distort the photon energy  
spectrum and could thus lead to a bias for the 
resulting $F_2$ contribution.  
The quality of the efficiency simulation has been checked 
using the photon energy spectrum of 
radiative ${\mathrm e^+ e^-}$ events where a high 
energy electron ($>43$ GeV) has been required to tag the event
in comparison with corresponding MC events.  
The efficiency ratio between data and MC is constant as
a function of $E_\gamma$ and consistent with unity to $\pm 6\%$.  
The resulting effect on the limit amounts to less than $\pm 0.0005$. 

{\it Background from non-$\tau$ events} has been estimated using 
the MC simulation considering all background reactions mentioned above
and is found to be very small ($0.13\%$). 
The total predicted background from ${\mathrm e^+ e^-}$, 
$\mu^+\mu^-$, and hadronic
processes amounts to $1.8\pm 1.8$ events.
The worst case 
assumption is that the background is distributed as the Standard Model 
expectation thereby artificially improving the limit. 
The resulting upper limit on $F_2$, when the background is included, is
however unchanged. 
 
Because the MC event sample is about $4.5$ times larger than the data 
we have neglected the {\it statistical error of the MC} in the fit. 
To check the validity of this assumption, 
the fit has also been performed using a method \cite{barlow} which
allows for the inclusion of both data and MC error in the likelihood.
The resulting limit changes negligibly 
by 0.0003 with respect to that obtained without
using the MC error.    

Assuming the systematic errors to be independent, they have been added
in quadrature. Then  
the total systematic uncertainty 
has been quadratically added to 
the statistical error in each bin of the $E_\gamma$ distribution
and a new limit has been derived 
\footnote{
Because the Poisson--based likelihood method 
does not have an explicit error term,
$\chi^2$ has been used for this estimation.}
as 
\begin{equation} 
|F_2|  < 0.067  \qquad {\rm at}  \qquad 95 \% \, {\rm C.L.} 
\end{equation}

\section*{Inclusion of the Interference Term} 
 
It has been shown \cite{riemann} for the cross section of
the process \eettg that the contribution of 
the interference between the standard part and the 
magnetic part of the electromagnetic current (see eq.\,(\ref{current}))
can be important even if $F_2$ is as large as several percent. 
The authors of \cite{riemann} have 
computed the differential cross section in the photon variables
$E_\gamma$ and $\cos \: \theta_\gamma$ for the \eettg 
process including the interference term. 
While this computation cannot serve for event
simulation by means of 4-vector generation, it has been used 
to study the relative importance of both terms.

\begin{figure}[htb] 
\centerline{\epsfig{file=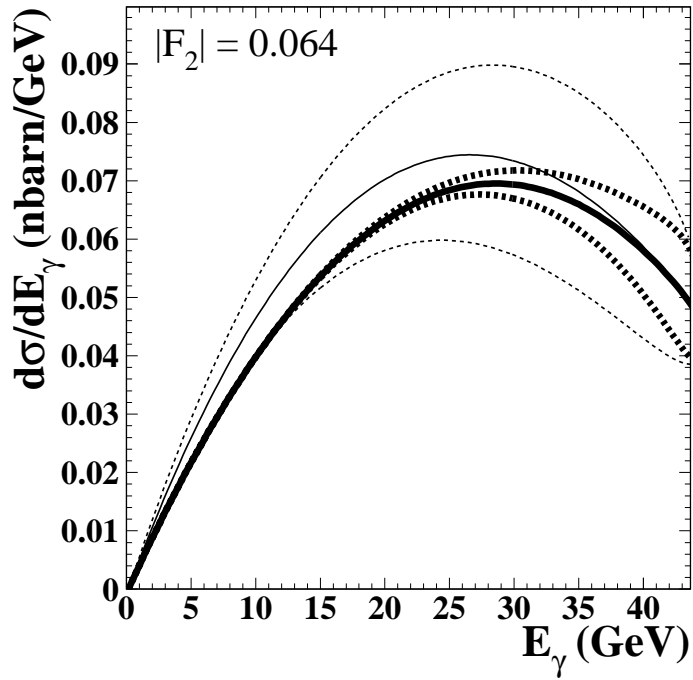,width=100mm}} 
{\caption{\label{interf}
{\it  Influence of the interference term; 
the thin lines show the cross-section without any additional cuts applied
(solid for the $|F_2|^2$ term alone and dashed with interference term included).
The thick lines (solid and dashed, respectively) show the result when a
minimum angle of $35^\circ$ between the photon and the $\tau$-cone axis
is required.}}}
\end{figure} 

Figure\,\ref{interf} shows the contribution of the interference term to the 
cross section assuming $|F_2| = 0.064$ for two different cuts on the event 
topology: without any requirement on the angular 
separation between the photon and the $\tau$ leptons (thin curves), 
and requiring a minimum angular separation of $35^\circ$ (thick curves)
as in this analysis. In each case, the central (solid) curve 
shows the cross section assuming an $|F_2|^2$ contribution only, 
while the upper/lower (dashed) curves are obtained by  
including the interference term with a positive/negative sign.
It is evident that the effect of the interference term is much reduced 
by requiring a minimum angular separation between the photon and the 
$\tau$ leptons. 
Nevertheless, the effect of  
the interference term can be taken into account to
obtain a limit on $F_2$ respecting its sign.
 
A correction of the signal spectrum 
is obtained by reweighting the signal $E_\gamma$ 
spectrum according to 
fig.\,\ref{interf} using different weighting factors for each 
value of $F_2$. 
Obviously, adding the interference term with 
a positive sign leads to a shift of 
the photon energy spectrum towards higher energies leading 
to an even better 
distinction between the SM and the $F_2$ spectrum, 
while for the negative sign the opposite is true.
This treatment assumes equal efficiencies for events due to the
interference term and to the quadratic term, an assumption which 
is not entirely correct. However, as long as the efficiency for the 
interference term is not larger than that of the quadratic term,
this assumption is conservative and is, in fact,
supported by the angular distributions shown in fig.\,\ref{signal}(c).
The interference term distribution must lie in between KORALZ and the
$|F_2|^2$ spectrum and due to the angular cuts its acceptance is
smaller acceptance than that of the $|F_2|^2$ term. 
One then obtains the following $95\%$ C.L. limit using the 
likelihood curve of fig.\,\ref{likeli}(b) and including systematic errors

\begin{eqnarray} 
-0.068 \, < \, F_2 \, < \, 0.065  \qquad .
\end{eqnarray}

\section*{Discussion and Conclusions} 
 
We have studied the reaction \eettg to search for a 
contribution from the anomalous   
magnetic form factor $F_2$ that is related to the anomalous magnetic moment  
$a_\tau$ of the $\tau$ lepton. 
The contribution of the $F_2$ form factor 
changes the distributions of the kinematic variables 
of the final state, most notably the photon energy spectrum.   
No significant contribution in addition to the Standard Model prediction 
is needed to describe the data. 
Comparing the data to the Standard Model prediction, a 95\% confidence level 
limit on the magnitude of the 
magnetic form factor $F_2$ of 
\begin{eqnarray} 
|F_2| \, < \, 0.067 
\end{eqnarray} 
has been placed. Taking into account the effect of the interference term
between the Standard Model amplitude and the $F_2$ amplitude 
the 95\% CL boundary on $F_2$ is 
\begin{eqnarray} 
-0.068 \, < \, F_2 \, < \, 0.065 \qquad . 
\end{eqnarray} 

\noindent 
Substituting $\displaystyle \frac{F_2}{2 m_\tau} 
\displaystyle \rightarrow \displaystyle \frac{F_3}{e} $
the bounds on $F_2$ translate to limits on $F_3$,
the electric dipole form factor of the $\tau$ lepton, for which
one obtains\footnote{A compilation of recent bounds on electric and weak dipole moments of the $\tau$ lepton can be found in \cite{PDG,nnw}.}  
 
\begin{equation} 
-3.8 \times 10^{-16}\, e \,{\rm cm} \, < \, eF_3 \,  
< \, 3.6 \times 10^{-16}\, e \, {\rm cm} \qquad ,
\end{equation} 

\noindent 
with the same interpretation 
restrictions as mentioned for $F_2$ in the introduction and neglecting   
a possible influence of the $\tau$ polarization on the term linear
in $F_3$.

\section*{Appendix} 

The formulae for the differential cross section \cite{ZEPP1} for
\eettg are given below using the $F_1$ (SM) and the $F_2$ terms
in the amplitude of eq.\,(\ref{amplitude}), but no interference.
 
$\left. \begin{array}{l} p_1 \:=\: \frac{\sqrt{s}}{2}\:(1,0,0, 1) \\ \\ p_2 \:=\: 
\frac{\sqrt{s}}{2}\:(1,0,0,-1) \end{array} \right\} \Rightarrow p_{\rm tot} \:
=\: p_1 + p_2 = \sqrt{s}\:(1,0,0,0)$ \\

$\left. q \:=\: E_\gamma \: (1,\sin\theta_\gamma \cos\phi_\gamma,\sin\theta_\gamma \sin\phi_\gamma,\cos\theta_\gamma) \right.$ 

$\left. k_1 \:=\: E_{\tau^-} \: (1,\hat{x}\sin\theta_{\tau^-\gamma} \cos\phi_{\tau^-} \:+\:\hat{y}\sin\theta_{\tau^-\gamma} \sin\phi_{\tau^-} \:+\: \hat{z}\cos\theta_{\tau^-\gamma}) \right.$ 

$\left. k_2 \:=\: p_{\rm tot} - q - k_1 \right.$ 

$\left. Q^\mu \:=\: \left(k_1 \cdot q \right)\:{k_2}^\mu \:-\: \left( k_1 \cdot k_2 \right)\:q^\mu \:+\: \left( k_2 \cdot q \right)\:{k_1}^\mu \right.$

\begin{eqnarray}
\hat{x} \:=\: \left( \begin{array}{c} -\cos\theta_\gamma \cos\phi_\gamma \\ -\cos\theta_\gamma \sin\phi_\gamma \\ \sin\theta_\gamma \end{array} \right) & \hat{y} \:=\: \left( \begin{array}{c} \sin\phi_\gamma \\ -\cos\phi_\gamma \\ 0 \end{array} \right) & \hat{z} \:=\: \hat{q} \:=\: \left( \begin{array}{c} \sin\theta_\gamma \cos\phi_\gamma \\ \sin\theta_\gamma \sin\phi_\gamma \\ \cos\theta_\gamma \end{array} \right) \nonumber
\end{eqnarray}

\begin{displaymath}
\frac{d \sigma \;(e^+ e^- \rightarrow \tau^+ \tau^- \gamma)}{dE_{\tau^-} dE_\gamma d\cos\theta_\gamma d\phi_{\tau^-} d\phi_\gamma} \quad=\quad 
\frac{\alpha^3}{2\pi^2 \sin^4\theta_W \cos^4\theta_W} \cdot \frac{1}{(s-{M_{{\rm Z}^0}}^2)^2 \:+\: (M_{{\rm Z}^0}\Gamma_{{\rm Z}^0})^2}
\end{displaymath}
\begin{displaymath}
\left\{ \left({c_v}^2+{c_a}^2\right)^2 \left[ \frac{2}{(k_1 q) (k_2 q)} \left[ (k_1 k_2) (p_1 p_2) - (k_1 p_1)(k_1 p_2) - (k_2 p_1)(k_2 p_2) \right] + \frac{(k_1 q)}{(k_2 q)} + \frac{(k_2 q)}{(k_1 q)} \right] \right.
\end{displaymath}
\begin{displaymath}
\left. + \; 4\: {c_v}^2 {c_a}^2 \left[ 2\: \frac{(k_1 p_2)(k_2 p_1)-(k_1 p_1)(k_2 p_2)}{(k_1 q)(k_2 q)} \:+\: q\: (p_2 - p_1) \left( \frac{1}{(k_1 q)} \:-\: \frac{1}{(k_2 q)} \right) \right] \right\}
\end{displaymath}
\begin{displaymath}
+\; {\bf{F_2}^2} \cdot \frac{\: \alpha^3}{\pi^2 \:{m_\tau}^2 \sin^4\theta_W \cos^4\theta_W} \cdot \frac{1}{s \left[ (s-{M_{{\rm Z}^0}}^2)^2 \:+\: (M_{{\rm Z}^0}\Gamma_{{\rm Z}^0})^2 \right]}
\end{displaymath}
\begin{displaymath}
\left\{ \left({c_a}^4-{c_v}^4\right) \left( \frac{(p_1 Q)(p_2 Q)}{(k_1 q)(k_2 q)} \:+\: (k_1 k_2)(p_1 p_2) \right) \right.
\end{displaymath}
\begin{displaymath}
+\; 2\: {c_v}^2 \left({c_v}^2+{c_a}^2\right)^2 \left[ (k_1 p_2)(k_2 p_1) \:+\: (k_1 p_1)(k_2 p_2) \right]
\end{displaymath}
\begin{equation}
\label{masterformel}
\Biggl. +\; 4\: {c_v}^2{c_a}^2 \left[ (k_1 p_2)(k_2 p_1) \:-\: (k_1 p_1)(k_2 p_2) \right] \Biggr\}
\end{equation}

\subsection*{Acknowledgments} 
This work has benefited much from numerous discussions with
Dieter Zeppenfeld who has also calculated the cross section
for the anomalous photon production. We would like to thank him
as well as J. Biebel and T. Riemann for providing the calculations
for the interference term. We would also like to acknowledge  
useful discussions with J. Swain, L. Taylor and O. Nachtmann.\\
Furthermore we
particularly wish to thank the SL Division for the efficient operation
of the LEP accelerator at all energies
and for
their continuing close cooperation with
our experimental group.  We thank our colleagues from CEA, DAPNIA/SPP,
CE-Saclay for their efforts over the years on the time-of-flight and trigger
systems which we continue to use.  In addition to the support staff at our own
institutions we are pleased to acknowledge the  \\
Department of Energy, USA, \\
National Science Foundation, USA, \\
Particle Physics and Astronomy Research Council, UK, \\
Natural Sciences and Engineering Research Council, Canada, \\
Israel Science Foundation, administered by the Israel
Academy of Science and Humanities, \\
Minerva Gesellschaft, \\
Benoziyo Center for High Energy Physics,\\
Japanese Ministry of Education, Science and Culture (the
Monbusho) and a grant under the Monbusho International
Science Research Program,\\
German Israeli Bi-national Science Foundation (GIF), \\
Bundesministerium f{\"u}r Bildung, Wissenschaft,
Forschung und Technologie, Germany, \\
National Research Council of Canada, \\
Research Corporation, USA,\\
Hungarian Foundation for Scientific Research, OTKA T-016660, 
T023793 and OTKA F-023259.\\

\end{document}